\documentstyle[12pt,epsf]{article}
\newcommand{\be}{\begin{equation}}
\newcommand{\ee}{\end{equation}}
\newcommand{\bea}{\begin{eqnarray}}
\newcommand{\eea}{\end{eqnarray}}
\newcommand{\nn}{\nonumber}

\newcommand{\NP}{Nucl. Phys.}

\newcommand{\PRL}{Phys. Rev. Lett.}
\newcommand{\PL}{Phys. Lett.}
\newcommand{\PR}{Phys. Rev.}
\def\art{\@ifnextchar[{\eart}{\oart}}
\def\eart[#1]#2#3#4#5#6{{\rm #2}, {\em #3 \rm #4} {\rm (#6) #5 ({\em #1})}}
\def\hepart[#1]#2{{\rm #2, \em#1}}
\newcommand{\oart}[5]{{\rm #1}, {\em #2 \rm #3} {\rm (#5) #4}}

\begin{document}

\def\gamh{\Gamma_H}
\def\eb{E_{\rm beam}}
\def\deb{\Delta E_{\rm beam}}
\def\sigm{\sigma_M}
\def\sigmmax{\sigma_M^{\rm max}}
\def\sigmmin{\sigma_M^{\rm min}}
\def\sige{\sigma_E}
\def\dsigm{\Delta\sigma_M}
\def\mh{M_H}
\def\lyear{L_{\rm year}}

\def\wstar{W^\star}
\def\zstar{Z^\star}
\def\ie{{\it i.e.}}
\def\etal{{\it et al.}}
\def\eg{{\it e.g.}}
\def\pzero{P^0}
\def\mt{m_t}
\def\mpzero{M_{\pzero}}
\def\mev{~{\rm MeV}}
\def\gev{~{\rm GeV}}
\def\gam{\gamma}
\def\lsim{\mathrel{\raise.3ex\hbox{$<$\kern-.75em\lower1ex\hbox{$\sim$}}}}
\def\gsim{\mathrel{\raise.3ex\hbox{$>$\kern-.75em\lower1ex\hbox{$\sim$}}}}
\def\ntc{N_{TC}}
\def\epem{e^+e^-}
\def\tauptaum{\tau^+\tau^-}
\def\lplm{\ell^+\ell^-}
\def\anti{\overline}
\def\mw{M_W}
\def\mz{M_Z}
\def\fbi{~{\rm fb}^{-1}}
\def\mupmum{\mu^+\mu^-}
\def\rts{\sqrt s}
\def\sigrts{\sigma_{\tiny\rts}^{}}
\def\sigrtssq{\sigma_{\tiny\rts}^2}
\def\sigrtsprime{\sigma_{E}}
\def\nsigrts{n_{\sigrts}}
\def\gampzero{\Gamma_{\pzero}}
\def\pzerop{P^{0\,\prime}}
\def\mpzerop{M_{\pzerop}}
\def\f{\frac}
\def\gt{\tilde g}
\def\ct{{\tilde c}_\theta }
\def\st{{\tilde s}_\theta }
\def\sb{s_\beta }
\def\cb{ c_\beta }
\def\eps{\epsilon}

\def\s{s_\theta }
\def\c{c_\theta }
\def\sz{\sum_{n=0}^\infty}
\def\su{\sum_{n=1}^\infty}
\font\fortssbx=cmssbx10 scaled \magstep2
\hbox to \hsize{
%
%
$\vcenter{
\hbox{\fortssbx University of Florence}
\hbox{\fortssbx University of Geneva}
}$
\hfill
$\vcenter{
\hbox{\bf DFF-341/7/99}
\hbox{\bf UGVA-DPT-1999 07-1047}
}$
}

%
\medskip
\begin{center}

{\Large\bf\boldmath SM Kaluza-Klein Excitations and
Electroweak Precision Tests\\}
\rm
\vskip1pc
{\Large R. Casalbuoni$^{a,b}$,  S. De Curtis$^b$,
\\ D. Dominici$^{a,b}$, R. Gatto$^c$\\}
\vspace{5mm}
{\it{$^a$Dipartimento di Fisica, Universit\`a di Firenze, I-50125
Firenze, Italia
\\
$^b$I.N.F.N., Sezione di Firenze, I-50125 Firenze, Italia\\
$^c$D\'epart. de Physique Th\'eorique, Universit\'e de
Gen\`eve, CH-1211 Gen\`eve 4, Suisse}}
\end{center}
\bigskip
\begin{abstract}
We consider a minimal extension to higher dimensions of the
Standard Model, having one compactified dimension, and we study
its experimental tests in terms of electroweak data. We discuss
tests from high-energy data at the $Z$-pole, and low-energy
tests, notably from atomic parity violation data. This
measurement combined with neutrino scattering data strongly
restricts the allowed region of the model parameters. Furthermore
this region is incompatible at 95\% CL with the restrictions
from high-energy experiments.  Of course a global fit to all data is
possible but the $\chi^2_{\rm min}$ for degree of freedom is
unpleasantly large.
\noindent
\end{abstract}
\newpage
\section{Introduction}
Since the original proposal by Kaluza \cite{kaluza} and by Klein
\cite{klein} the possibility of compactified extra dimensions has
been a recurrent theme in theoretical physics, in particular in
supergravity and string theories. Recent developments have
suggested the possibility of low compactification scales
\cite{antoniadis,arkani}. There  have also been similar
developments on extensions of the Standard Model (SM) to include
compactified extra spatial dimensions \cite{quiros}. A simplest
minimal extension is to 5 dimensions, leading to towers of
Kaluza-Klein (KK) excitations of the SM gauge bosons. Within such
an extension, precision electroweak measurements have been
analyzed to derive bounds on the compactification scale
\cite{rizzo,strumia}.

In this note we present a new discussion of these tests,
analyzing the roles played by precision high-energy measurements
at the $Z$-pole, as summarized in the $\epsilon$ parameters
\cite{altarelli} (or $S$, $T$, $U$ parameters \cite{peskin}), and
by low-energy experiments, in particular the new experiment on
the  weak charge $Q_W$ in atomic cesium \cite{bennett}, whose
implications for various models have already been analyzed in
\cite{decurtis}. The reason why we have chosen to describe the
high-energy measurements in terms of the $\eps$'s is that they
are just an efficient way of collecting all the relevant data
about the observables  and also because they can be expressed
easily in analytical way. This can be done by eliminating the KK
excitations and constructing an effective lagrangian around the
$Z$-pole and an effective four-fermi interaction (eliminating
also the $W$ and the $Z$) in the low-energy limit $E<<m_Z$. This
allows us also to perform a simple discussion of the low-energy
observables.

In this paper we  are mainly interested in studying the
compatibility of the model with all the available data. This is
done by considering separately the effects of the different
measurements on the parameter space. The reason why we choose
this procedure is because the low-energy experiments test
phenomenological aspects which cannot be tested from measurements
at the $Z$-pole. The low-energy experiments provide for
measurements of the effective electroweak coupling of the light
quarks and leptons, which in "new physics" schemes might not be
directly related to the studies done at the $Z$-pole.

In particular we have  performed a fit to the high-energy data,
by obtaining bounds on the compactification scale in terms of the
mixing of the KK excitations with the $Z$. These data alone leave
room to the  new physics described here, in fact the $\chi^2_{\rm
min}$/d.o.f. of the fit turns out to be acceptable. On the other
hand the low-energy data leave almost no space to the model. This
is mainly due to the new data on Atomic Parity Violation (APV)
which put a lower limit on the mixing angle $\beta$  of this
model allowing only the region around the maximal mixing. In fact
the new results on APV, if taken literally, disfavor the SM and
all the models giving rise to negative extra contributions to
$Q_W$ at more than 99\% CL. In the maximal mixing region, where
the extra contribution to $Q_W$ turns out to be positive, the
neutrino scattering experiments as NuTeV and CHARM cut off most
of the region allowed by $Q_W$. Furthermore, the regions allowed
by $Q_W$ and by the high-energy data result to be incompatible at
95\% CL.

One could also combine all the previous data obtaining bounds on
the parameter space. We have tried to do this by putting together
the high-energy data with the $Q_W$ measurement. The result is
that, although the allowed region is not very much dissimilar to
the one obtained using only the high-energy data, the
$\chi^2_{\rm min}$/d.o.f. deteriorates in a considerable way.

In Section 2 we present the effective lagrangian obtained after
eliminating the KK modes. This lagrangian is useful to describe
physics at energies much lower than the compactification scale,
as for $E\approx m_Z$. In Sections 3 and 4 we discuss the bounds
from the high-energy and the low-energy data respectively.
Conclusions are given in Section 5.

\section{An extension of the SM in 5 dimensions}
 The extra dimension model we consider, is
the one suggested by \cite{quiros} and it is based on an
extension of the SM to 5 dimensions. The fifth dimension $x^5$ is
compactified on a circle of radius $R$ with the identification
$x^5\to -x^5$ ($Z_2$-parity). Gauge fields and one Higgs field
($\phi_1$) live in the bulk, fermions and a second Higgs field
$\phi_2$ live on the 4D wall (boundary). The action for the SM
with one extra dimension  is obtained from the action for the
general 5D gauge theory. The fields living in the bulk are
defined to be even under the $Z_2$-parity. In the case of an
$SU(2)_L\otimes U(1)$ gauge group, integrating over the fifth
dimension, the resulting  4D theory (in the unitary gauge) is
given by (omitting  the kinetic terms):
\bea
{\cal L}^{charged}&=&\f 1 8 \gt^2 v^2 \Big[ W_1^2+\cb^2 \su
(W_1^{(n)})^2 + 2 \sqrt {2}
\sb^2 W_1 \su  W_1^{(n)}\nn\\
&+& 2 \sb^2  ( \su W_1^{(n)})^2\Big] + \f 1 2 \su n^2 M^2
(W_1^{(n)})^2 \nn\\ &-& \gt ( W_1^\mu +\sqrt{2}
\su W_1^{(n)\mu}) J_\mu^1 +( 1\to 2)
\label{charged}
\eea

\bea
{\cal L}^{neutral}&=&\f 1 8 \f {\gt^2}{\ct^2} v^2
\Big[ Z^2+\cb^2 \su (Z^{(n)})^2 + 2 \sqrt {2}
\sb^2 Z \su  Z^{(n)}\nn\\
&+& 2 \sb^2  ( \su Z^{(n)})^2\Big] + \f 1 2 \su n^2 M^2
[(Z^{(n)})^2 +(A^{(n)})^2]\nn\\ &-&\f {\tilde e }{\st\ct} (Z^\mu
+\sqrt{2} \su Z^{(n)\mu}) J_\mu^Z - {\tilde e} (A^\mu
 +\sqrt{2} \su A^{(n)\mu}) J^{em}_\mu
\label{neutral}
\eea
where  $M=1/R$,  $\gt$ and $\gt'$ are the gauge couplings and
$\tan\tilde\theta=
\st/\ct=\gt'/\gt$, $\tilde e=\gt\st$. $W^{(n)}$, $Z^{(n)}$ and
$A^{(n)}$ are the KK excitations of the standard $W$, $Z$ and $A$
fields, $<\phi_{1}>=v\cos\beta\equiv v\cb$,
$<\phi_{2}>=v\sin\beta\equiv v\sb$, and
\bea
J_\mu^{1,2}&=&\bar \psi_L \gamma_\mu \f {\tau_{1,2}}
{2}\psi_L\nn\\ J_\mu^Z&=&\bar \psi \gamma_\mu (g_v+\gamma_5
g_a)\psi,~~~~g_v=\f {T_{3L}} 2- \st^2 Q, ~~~~g_a=-\f {T_{3L}} 2
\nn\\J_\mu^{em}&=&\bar \psi  \gamma_\mu Q\psi,~~~~
Q = T_{3L} +\f{B-L} 2
\label{j1}
\eea
The effects of the KK excitations in the low-energy limit
$p^2<<M^2$, can be studied by eliminating the corresponding
fields using the solutions of their equations of motion for
$M\to\infty$ \cite{anic}. In this limit the kinetic terms are
negligible and one gets:
\bea
W^{(n)}_{1,2}&\sim& - \sqrt{2}\f {\sb^2}{n^2}\f{{\tilde
m}^2_W}{M^2} W_{1,2}+\sqrt{2}\f {\gt}{n^2 M^2} J^{1,2}\nn\\
Z^{(n)}&\sim&-
\sqrt{2}\f {\sb^2}{n^2}\f{{\tilde m}^2_Z}{M^2} Z+\sqrt{2}\f {\tilde e}{\st\ct}
\f 1{n^2M^2}J^Z\nn\\
A^{(n)}&\sim&\sqrt{2} \f {\tilde e}{n^2 M^2} J^{em}
\label{soluz}
\eea
where ${\tilde m}^2_W= \gt^2 v^2/4$ and  ${\tilde m}^2_Z={\tilde
m}^2_W/\ct^2$. We use tilded quantities to indicate that, due to
the effects of the KK excitations, they are not the physical
parameters. We do not use tilded notations for the fields because
they are not renormalized at the order ${\cal
O}(1/M^2)$.

By using eq. (\ref{soluz}) in eqs. (\ref{charged}) and
(\ref{neutral}), one can read the mass values for $W$ and $Z$ (to
order ${\cal O}(1/M^2)$)
\be
m^2_W = {\tilde m}_W^2 [1-\ct^2\sb^4 X],~~~~ m^2_Z ={\tilde
m}_Z^2 [1- \sb^4 X]
\label{masse}
\ee
where
\be
X=\f {\pi^2} {3} \f {m_Z^2}{M^2}
\ee

The effective couplings are given by
\be
{\cal L}^{charged}_{eff} = -\gt J^1_\mu W^{1\mu} (1-\sb^2 \ct^2
X)-\f{\gt^2}{2 m_Z^2} X J^1_\mu J^{1\mu}+(1\to 2)
\label{effch}
\ee

\bea
{\cal L}^{neutral}_{eff} &=& -\f e{\st\ct} J^Z_\mu Z^{\mu}
(1-\sb^2
 X)-\f{e^2}{2 \st^2\ct^2  m_Z^2} X J^Z_\mu J^{Z\mu}\nn\\ &-&e
 J^{em}_\mu A^{\mu}-\f{e^2}{2  m_Z^2}  X J^{em}_\mu J^{em \mu}
 \label{effneu}
\eea
The electric charge $e$ is identified as  $e=\tilde e =\gt \st$
since, being  defined at zero momentum, it is not renormalized.
Notice the presence  in the effective lagrangian  of four fermion
interactions which give additional  contributions to the Fermi
constant and to neutral current processes.

\section{Bounds from the high-energy data}
Let us now study the low-energy effects ($E<<M$) of the KK
excitations. By using the effective lagrangian (\ref{effch}) we
can evaluate the Fermi constant
\be
G_F= \f {\sqrt{2}\gt^2}{8 m_W^2}[1+\ct^2 X][1-2 \sb^2 \ct^2 X]
\label{gf}
\ee
Using eq. (\ref{masse}), one gets
\be
\f {G_F}{\sqrt{2}}= \f {e^2}{8 \st^2\ct^2 m_Z^2} (1 +\Delta)
\ee
where
\be
\Delta = \ct^2 X ( 1 - 2 \sb^2 - \sb^4 \f {\st^2} {\ct^2})
\label{delta}
\ee
By defining an effective $\theta$ angle by $ {G_F}/\sqrt{2}=
 {e^2}/({8 \s^2\c^2 m_Z^2})$
we have
\be
\s^2 = \st^2 (1- \f {\c^2 }{c_{2\theta}}\Delta),~~~~
\c^2 = \ct^2 (1+\f {\s^2 }{c_{2\theta}}\Delta)
\label{st}
\ee
By defining
\be
\f {m_W^2}{m_Z^2} = \c^2 (1-\f {\s^2 }{c_{2\theta}}
\Delta r_W)
\ee
we get
\be
\Delta r_W = \Delta -  {c_{2\theta}} \sb^4 X
= \c^2 X(1- 2 \sb^2 - \sb^4)
\ee
Furthermore using the neutral couplings  of $Z$ to fermions
of eq. (\ref{effneu}) and using the definitions of $\Delta\rho$
and $\Delta k$ given in  \cite{altarelli}
\bea
&&-2(\sqrt{2}G_F)^{1/2}m_Z (1+\f 1 2 \Delta\rho )J^{Z}_\mu Z^\mu\nn\\
&&
J_\mu^Z=\bar \psi\gamma_\mu (g_v+\gamma_5
g_a)\psi\nn\\
&&g_v=\f {T_{3L}} 2- \s^2(1+\Delta k) Q, ~~~~g_a=-\f {T_{3L}} 2
\label{jz}
\eea
we find
\be
\Delta\rho = - \Delta - 2  {\sb^2} X
=-\c^2 X [1 + \f {\s^2}{\c^2} \sb^2 (1+ \cb^2)]
\ee
\be
\Delta k = \f {\c^2}{c_{2\theta}} \Delta
= \f {\c^4}{c_{2\theta}} X (1 - 2 \sb^2 - \sb^4 \f {\s^2} {\c^2})
\ee
Then we can easily compute the contribution to the $\eps$
parameters \cite{altarelli} coming from new physics
\bea
\eps_{1N}&=& -\c^2 X [ 1 + \sb^2 \f {\s^2}{\c^2} (1+ \cb^2)]\nn\\
\eps_{2N}&=& -\c^2 X\nn\\
\eps_{3N}&=& - 2 \c^2 \sb^2 X\label{eps}
\eea
We have studied the bounds on the model by considering the
experimental values of the $\epsilon$ parameters coming from all
the high-energy data \cite{altadati}
\bea
\eps_1&=& (3.92\pm 1.14)\cdot 10^{-3}\nn\\
\eps_2&=& (-9.27\pm 1.49)\cdot 10^{-3}\nn\\
\eps_3&=& (4.19\pm 1.00)\cdot 10^{-3}
\label{epsd}
\eea
We have added to eq. (\ref{eps}) the contribution from the
radiative corrections, assuming  that they are the same as in the
SM. For $m_t=175~GeV$ and $m_H=100(300)~GeV$ one has
\cite{altarelli2}: $\eps_1^{\rm rad}= 5.62(4.97)\cdot 10^{-3}$,
$\eps_2^{\rm rad}= -7.54(-7.18)\cdot 10^{-3}$, $\eps_3^{\rm
rad}=5.11(6.115)\cdot 10^{-3}$. Notice that in principle there
could be also new contributions to the radiative corrections from
the additional charged and neutral Higgs bosons.

The $95\%$ CL lower bounds  on the scale $M$ at fixed $\sb$,
coming from the three $\eps$ observables,  are shown in Fig. 1
for  $m_H$ = 100 $GeV$ and $m_H$ = 300 $GeV$. The limit does not
depend strongly on $\sb$ and it is more restrictive for $\sb=1$,
where one gets a lower limit for the compactification scale
$M\sim 3.5~TeV$. Comparable results performing a global fit to
the electroweak precision observables have been obtained also by
\cite{quiros,rizzo,strumia}. These bounds are more stringent than
the ones from the Tevatron upgrade \cite{rizzo}.

\section{Bounds from  low-energy observables}
In this Section  we will be concerned with the low-energy
phenomenology ($E<<m_Z$). To this end we can also eliminate the
$W$ and $Z$ fields obtaining   effective current-current
interactions. In particular we will consider observables from
neutrino scattering and APV for which the relevant effective
lagrangian is
\be
{\cal L}^{{\rm low-en}}_{eff} = -4 \f{G_F}{\sqrt{2}} \Big[ (
J_\mu^1J^{\mu 1} +(1\to 2)) +(1+\s^2(1-\sb^2)^2 X) J_\mu^ZJ^{\mu
Z}\Big]
\label{le}
\ee
with $J_\mu^1$ and $J_\mu^Z$ given in eq. (\ref{j1}). Remember
that $J_\mu^Z$ contains a further correction from new physics
since it depends on $\st^2$. As an example we show the expression
for the atomic weak charge $Q_W$
\be
Q_W= {\bar Q}_{W} [1 + \s^2 X (\sb^2-1)^2]-4\frac{\s^2\c^2}
{c_{2\theta}}Z\Delta
\label{qwkk}
\ee
where ${\bar Q}_W$ has the SM expression for $Q_W$ evaluated at
$\s$ given in eq. (\ref{st}), $Z$ is the atomic number and
$\Delta$ is given in eq. (\ref{delta}). In a recent paper
\cite{bennett} a new determination of the weak charge of the
atomic cesium has been reported. This takes advantage from a new
measurement of the tensor transition probability for the $6S\to
7S$ transition in cesium and from improving the atomic structure
calculations in light of new experimental tests. The value
reported  $Q_W(^{133}_{55}Cs)=-72.06\pm (0.28)_{\rm expt}\pm
(0.34)_{\rm theor}$ represents a big improvement with respect to
the old determination \cite{noecker,blundell} since it gives a
measure of $Q_W$ at a level ($\approx
.6 \%$) comparable with the results obtained in high-energy
experiments. On the theoretical side, $Q_W$ can be expressed in
terms of the $S$ parameter \cite{marciano}, or in terms of the
 $\eps_3$ parameter \cite{alta} (the dependence on the $T$, or
 $\eps_1$, parameters is negligible)
\be
Q_W=-72.72\pm 0.13-102\epsilon_3^{\rm rad}+\delta_NQ_W
\label{qwth}
\ee
including hadronic-loop uncertainty. In the above definition of
$Q_W$ we have included only SM contributions to radiative
corrections. New physics is represented by $\delta_N Q_W$. The
discrepancy between the SM and the experimental data is given by
\be
Q_W^{expt}-Q_W^{SM}=1.18(1.28)\pm 0.46~~~~{\rm for}~~~~
m_H=100(300)~GeV
\label{qwexpt}
\ee
Therefore if we believe that the latest experiment and the
theoretical determinations of the atomic structure effects are
correct, we conclude that a negative or zero new physics
contribution  is excluded at more than 99\% CL.

In Fig. 2 we show the $95\%$ CL bounds on  $M$ at fixed $\sb$
from $Q_W$ (thick  solid lines). The allowed region is the
shadowed one. The actual $Q_W$ measurement leads to a lower bound
$\sb>0.707$. The reason why the region  $\sb<0.707$ is excluded at
95\% CL by the new APV measurement
is that the new physics contribution to $Q_W$ in that
region is negative definite.

In the same figure we have also plotted the 95\% CL bounds coming
from other low-energy observables. Namely we have considered the
following list of neutrino-nucleon scattering processes
\cite{zeppe,langa}:
\begin{itemize}
\item NuTeV experiment which measures a ratio of cross-sections of
neutral and charged currents denoted by $R^-$ \cite{nutev}. The
experimental result is $R^-=0.2277\pm 0.0021\pm 0.0007$, whereas
the SM value is $R^-_{SM}=0.2297\pm0.0003$;
\item CCFR experiment which
measures a different ratio of neutral and charged cross-sections
with the result \cite{ccfr} $R_\nu=0.5820\pm 0.0027\pm 0.0031$ to
be compared with the SM value $R_\nu^{SM}=0.5827\pm 0.0005$;
\item The experiments CHARM \cite{charm} and CDHS \cite{cdhs} which
give the results
\bea
R_\nu&=& 0.3021\pm 0.0031\pm 0.0026,~~~~{\rm CHARM}\nn\\ R_\nu&=&
0.3096\pm 0.0033\pm 0.0028,~~~~{\rm CDHS}\nn\\ R_{\bar\nu}&=&
0.403\pm 0.014\pm 0.007,~~~~{\rm CHARM}\nn\\ R_{\bar\nu}&=&
0.384\pm 0.016\pm 0.007,~~~~{\rm CDHS}\nn
\eea
with the SM values given by $R_\nu^{SM}=0.3089\pm 0.0003$,
$R_{\bar\nu}^{SM}=0.3859\pm 0.0003$.
\end{itemize}
The expressions for all these observables  can be easily obtained
from the effective lagrangian in eq. (\ref{le}), and therefore we
can evaluate the 95\% CL bounds on $M$ at fixed $\sb$. Contrarily
to $Q_W$, all these measurements are compatible with the SM
predictions and therefore the allowed regions in  Fig. 2 coming
from neutrino scattering experiments all contain the line $X=0$
($M\to\infty$), and therefore they lie in the upper part of the
corresponding lines.
We notice that the CHARM measurement of $R_\nu$ (thin  solid
line) and NuTeV result on $R^-$ (long dash line) practically cut
off  the whole region allowed by APV. One understands how this
happens by looking at the values of the pulls of $R_\nu$ and of
$R^-$, which are both negative (-0.9 and
-1.7 respectively \cite{langa}), whereas $Q_W$ has a positive pull
(2.6 \cite{bennett}). The model considered here provides for
positive deviations to $Q_W$, $R_\nu$, $R^-$ around
$\sin\beta=1$.

In order to compare the bounds from APV with the ones from the
high-energy precision measurements expressed in terms of the
$\eps$ parameters, we have  also drawn the latter ones in Fig. 2.
Notice that the two regions  are not compatible at 95\% CL. This
follows from  eqs. (\ref{qwth}) and (\ref{qwexpt}) which give a
95\% CL positive lower bound on new physics contributions to
$Q_W$ leading to an upper bound on the compactification scale
\cite{decurtis}. The same comparison is made in Fig. 1 for two
values of the Higgs mass ($m_H=100,~300~GeV$), showing that
increasing the Higgs mass the incompatibility between the two
regions remains.

Nevertheless if we consider a global fit to all the four
observables $\eps_1$, $\eps_2$, $\eps_3$ and $Q_W$, we have
bounds which are similar to the ones obtained by combining the
three $\eps$ variables shown in Fig. 1. However the fit  gives,
for $m_H=100~GeV$, $\chi^2_{\rm min}/{\rm d.o.f.}$ between 3 and
2.3 and, for $m_H=300~GeV$, $\chi^2_{\rm min}/{\rm d.o.f.}$
between 4.3 and 2.3 when varying $\sb$ . This must be compared
with the fit including only the $\eps$ parameters which gives a
$\chi^2_{\rm min}/{\rm d.o.f.}$ less than 1 for any $\sb$ at
$m_H=100~GeV$, and a   $\chi^2_{\rm min}/{\rm d.o.f.}$ less than
1 except for $\sb\approx 0$, where it is of order 2, for
$m_H=300~GeV$.

\section{Conclusions}

In this paper we  have tested  a minimal extension of the SM with
one extra spatial dimension against the existing data on
electroweak observables. In this analysis a relevant role is
played by the new results on APV. In fact, these data put a 95\%
CL lower limit on the mixing angle of the KK modes with the SM
gauge bosons allowing only the region around the maximal mixing
($\sin\beta>0.707$). The inclusion of other low-energy
measurements from neutrino scattering, further restricts this
region. In particular the data from CHARM and NuTeV practically
cut out all the allowed region. Also the high-energy precision
measurements put strong constraints on the compactification scale
which turn out to be incompatible with the restrictions from
$Q_W$. In fact the large deviations  from the SM required from
APV could be explained only with a compactification scale smaller
than the lower bounds from high-energy. For instance, we can read
from Fig. 1 that at $\sin\beta=1$, APV requires $M<2.6~TeV$,
whereas high-energy requires $M>3.6~TeV$.

Therefore our conclusion  is that, by  seriously taking the APV
result, the model  considered here is disfavored at 95\% CL  from
the actual measurements. We have also discussed the possibility
of performing a global fit to the data, by putting together the
high-energy and the APV measurements. In this case the region
allowed is practically coincident with the one coming from
high-energy (since these are the data with smaller error) but the
value of $\chi^2_{\rm min}$/d.o.f. turns out to be unpleasantly
large.

Finally let us notice that the dependence on the extra dimension
is hidden in the variable $X$ which parameterizes all the
deviations from the SM. In the case of more than one extra
dimension, one has $X\to 2\sum_{\vec n} F(\vec n)/(M^2 \vec
n^2)$, with $F(\vec n)$ a positive form factor, and $\vec n$ a
vector with positive integer components in the extra dimensions
\cite{rizzo}.
Since $X$ remains a positive definite quantity all our
conclusions about the incompatibility of APV and high-energy data
remain true. Only the numerical values of the bounds in the
compactification scale obtained for each observable depend on the
actual number of extra dimensions.
\medskip
\begin{center}
{\bf Acknowledgements}
\end{center}
\medskip
R.C.  would like to thank the Theory Division of CERN for the
kind hospitality offered to him during the final stage of this
paper.

\begin{figure}
\epsfysize=14truecm
\centerline{\epsffile{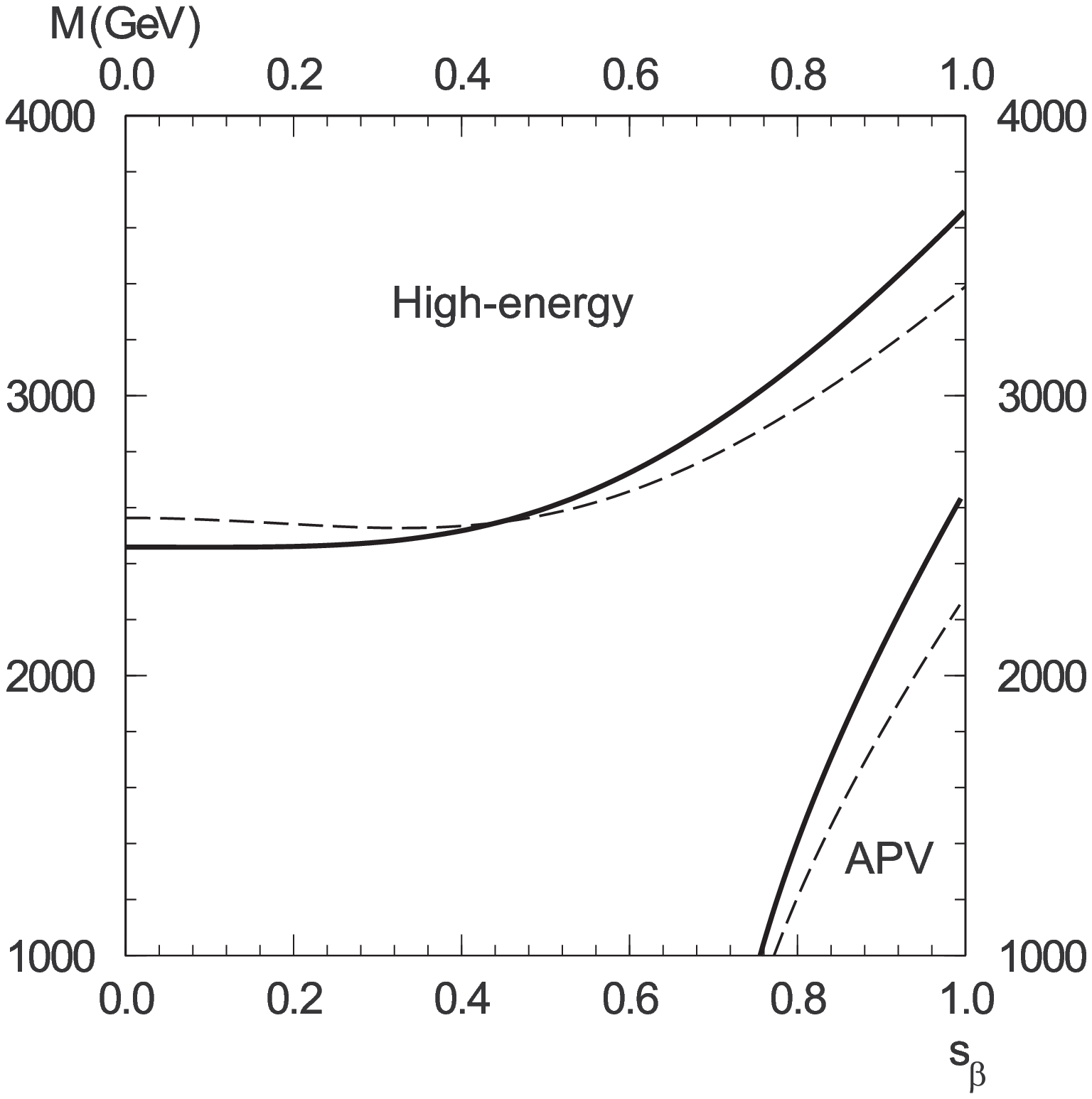}}
\noindent
\caption{
The 95\% CL  bounds on the compactification scale $M$ from  the
high-energy precision measurements ($\epsilon$ parameters)
and from APV measurement. The solid and the dash lines  correspond to
$m_H=100~GeV$ and $m_H=300~GeV$
respectively.}
\label{fig1}
\end{figure}

\begin{figure}
\epsfysize=13truecm
\centerline{\epsffile{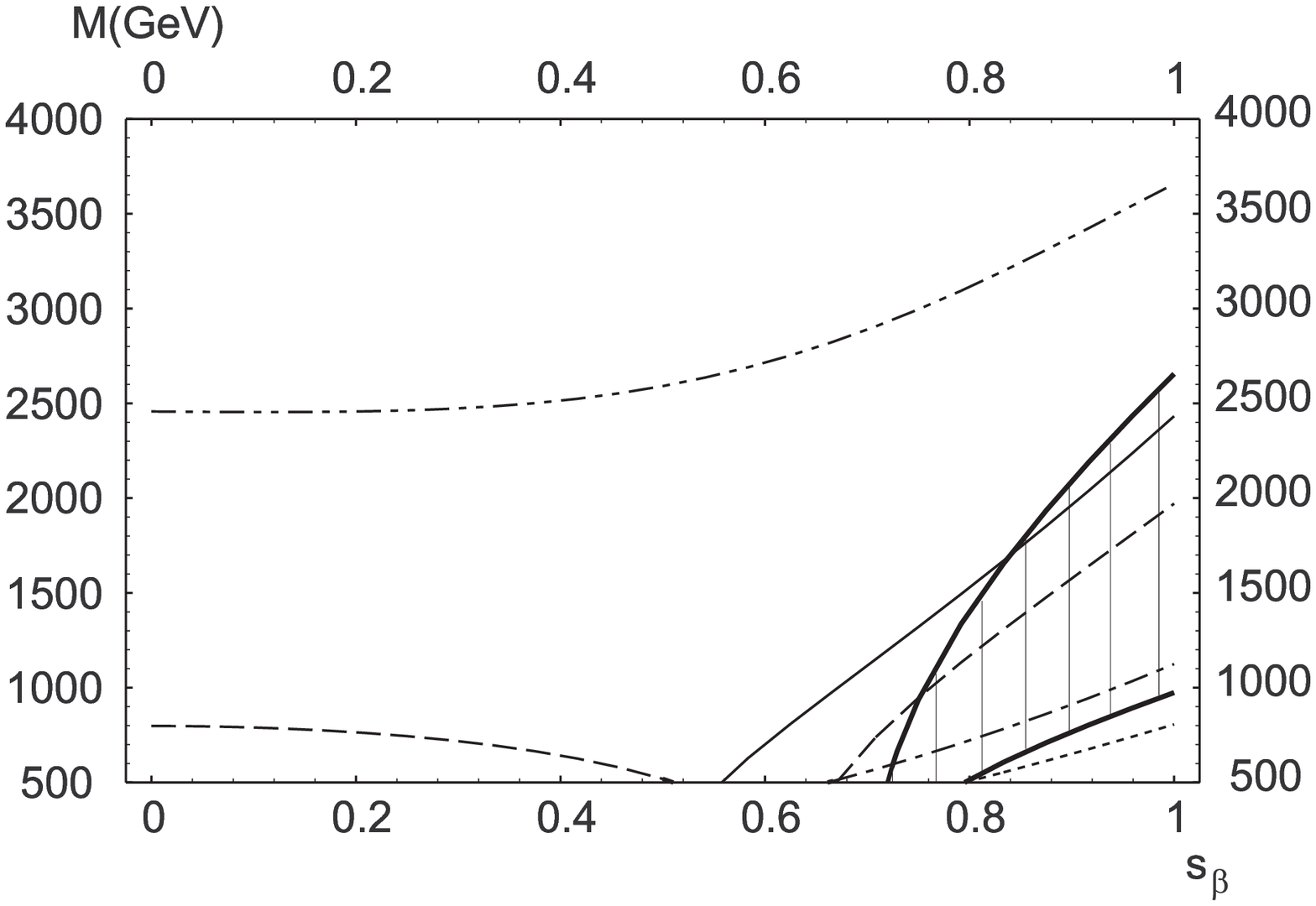}}
\noindent
\caption{
The 95\% CL  bounds on the compactification scale $M$ from
the low-energy experiments:
thick solid lines from $Q_W$ (the allowed region is the shadowed
one); thin solid line from CHARM measurement of $R_\nu$; long
dash lines from NuTeV; short dash line from CDHS measurement of
$R_\nu$; dash-dot line from CCFR. The measurement of
$R_{\bar\nu}$ gives  no restrictions on the plane shown in the
figure. For comparison we give also the bound from high-energy
measurements, dash-dot-dot line. The SM values of the observables have
been evaluated for a light Higgs mass.}
\label{fig2}
\end{figure}


\begin{thebibliography}{99}
\bibitem{kaluza}
T. Kaluza, Preuss. Akad. Wiss. (1921) 966.

\bibitem{klein}
O. Klein, Z. Phys. {\bf 37} (1926) 895.

\bibitem{antoniadis}
For some recent literature, see for instance: I. Antoniadis,
{\PL} {\bf B246} (1990) 377; E. Witten, {\NP} {\bf B471} (1996)
135; J.D. Lykken, {\PR} {\bf D54} (1996) 3693.

\bibitem{arkani}

N. Arkani-Hamed, S. Dimopoulos and G. Dvali, {\PL} {\bf B429}
(1998) 263; I. Antoniadis, N. Arkani-Hamed, S. Dimopoulos and G.
Dvali, {\PL} {\bf B436} (1998) 257; N. Arkani-Hamed, S.
Dimopoulos and G. Dvali, {\PR} {\bf D59} (1999) 086004; K.R.
Dienes, E. Dudas and T. Gherghetta {\PL} {\bf B43}(1998) 55.


\bibitem{quiros} A. Pomarol and M. Quiros, Phys. Lett. {\bf B438} 255 (1998);
A. Delgado, A. Pomarol and M. Quiros, hep-ph/9812489; M. Masip
and  A. Pomarol, hep-ph/9902467.

\bibitem{rizzo}
T.G. Rizzo and J.D. Wells, hep-ph/9906234; I. Antoniadis, K.
Benakli, M. Quiros,  Phys.Lett. {\bf B331} (1994) 313 ; {\it
ibidem} hep-ph/9905311.

\bibitem{strumia}
P. Nath and M. Yamaguchi, hep-ph/9902323; P. Nath and M.
Yamaguchi, hep-ph/9903298, W.J. Marciano, hep-ph/9903451; A.
Strumia, hep-ph/9906266.

\bibitem{altarelli}
G. Altarelli, R. Barbieri and S. Jadach, Nucl. Phys. {\bf B369}
(1992) 3; G. Altarelli, R. Barbieri and F. Caravaglios Nucl.
Phys. {\bf B405} (1993) 3; {\it ibidem} Phys. Lett. {\bf B349}
(1995) 145.

\bibitem{peskin}
M.E. Peskin and T. Takeuchi, {\PRL} {\bf 65} (1990) 964; {\it
ibidem} {\PR} {\bf D46} (1991) 381.

\bibitem{bennett}
S.C. Bennett and C.E. Wieman, Phys. Rev. Lett. {\bf 82} (1999)
2484.

\bibitem{decurtis}
R. Casalbuoni, S. De Curtis, D. Dominici and R. Gatto,
hep-ph/9905568.

\bibitem{anic} C.P. Burgess, S. Godfrey, H. Konig, D. London, I. Maksymyk,
 Phys. Rev. {\bf D49} (1994) 6115; L. Anichini, R. Casalbuoni and
 S. De Curtis, Phys. Lett, {\bf B348} (1995) 521.



\bibitem{altadati} G. Altarelli, presented at the Blois Conference,
June 1999.

\bibitem{altarelli2}
G. Altarelli, R. Barbieri and F. Caravaglios, Int. J. Mod. Phys.
{\bf A13} (1998) 1031.



\bibitem{noecker}
M.C. Noecker, B.P. Masterson and C.E. Wieman, Phys. Rev. Lett.
{\bf 61} (1988) 310.

\bibitem{blundell}
S.A. Blundell, W.R. Johnson and J. Sapirstein, Phys. Rev. Lett.
{\bf 65} (1990) 1411; V. Dzuba, V. Flambaum, P. Silvestrov and O.
Sushkov, Phys. Lett. {\bf A141} (1989) 147.

\bibitem{marciano}
W.J. Marciano and J.L. Rosner, Phys. Rev. Lett. {\bf 65} (1990)
2963.

\bibitem{alta} See the second reference in \cite{altarelli}.



\bibitem{zeppe}
For  recent reviews of the subject see: D. Zeppenfeld and K.
Cheung, hep-ph/9810277.

\bibitem{langa}
 J. Erler and P. Langacker, hep-ph/9903476.

\bibitem{nutev}
K.S. McFarland et al., hep-ex/9806013.

\bibitem{ccfr}
K.S. McFarland et al., Eur. Phys. J. {\bf C1} (1998) 509.

\bibitem{charm}
V. Allaby et al., Z. Physics {\bf C36} (1987) 611.

\bibitem{cdhs}
A. Blondel et al., Z. Physics {\bf C45} (1990) 361.



\end{thebibliography}
\end{document}